\documentclass[aps,prl,twocolumn,superscriptaddress]{revtex4-1}
\usepackage[urlcolor=blue, hyperindex, colorlinks, bookmarks=true]{hyperref}
\usepackage{graphicx}
\usepackage{times}
\usepackage{amsmath}
\usepackage{braket}
\usepackage{xcolor}
\usepackage{tabularx}
\usepackage[symbol]{footmisc}

\renewcommand{\eqref}[1]{Eq.~(\ref{#1})}

\newcommand{\removedD}[1]{{\color{gray}{#1}}}
\renewcommand{\removedD}[1]{{}} 

\newcommand{\corrected}[1]{}

\renewcommand{\eqref}[1]{Eq.~(\ref{#1})}

\newcommand{\appref}[1]{\hyperref[#1]{Appendix~\ref*{#1}}}
\newcommand{\tabref}[1]{\hyperref[#1]{Table~\ref*{#1}}}

\usepackage[english]{babel}

\begin{document}

\title{Single-Spin Relaxation in a Synthetic Spin-Orbit Field} 
\author{F.~Borjans}
\author{D.~M.~Zajac}
\author{T.~M.~Hazard}
\author{J.~R.~Petta}
\email{petta@princeton.edu}
\affiliation{Department of Physics, Princeton University, Princeton, New Jersey 08544, USA}
\date{\today}

\begin{abstract}
{
Strong magnetic field gradients can produce a synthetic spin-orbit interaction that allows for high fidelity electrical control of single electron spins. We investigate how a field gradient impacts the spin relaxation time $T_1$ by measuring $T_1$ as a function of magnetic field $B$ in silicon. The interplay of charge noise, magnetic field gradients, phonons, and conduction band valleys leads to a maximum relaxation time of $160\, \rm ms$ at low field, a strong spin-valley relaxation hotspot at intermediate fields, and a $B^4$ scaling at high fields. $T_1$ is found to decrease with lattice temperature $T_{\rm lat}$ as well as with added electrical noise. In comparison, samples without micromagnets have a significantly longer $T_1$. Optimization of the micromagnet design, combined with reductions in charge noise and electron temperature, may further extend $T_1$ in devices with large magnetic field gradients.
}
\end{abstract}

\maketitle

The Zeeman-split spin states of a single electron spin in a large magnetic field can naturally be used to define a qubit \cite{loss_quantum_1998,hanson_spins_2007}. For spins trapped in semiconductor quantum dots, single qubit rotations can be achieved through conventional electron spin resonance (ESR) with oscillating magnetic fields \cite{koppens_driven_2006,veldhorst_addressable_2014}. Spin rotations can also be electrically driven in the presence of an intrinsic spin-orbit field \cite{nowack_coherent_2007,nadj-perge_spinorbit_2010,petersson_circuit_2012} or a synthetic spin-orbit field generated by a micromagnet \cite{pioro-ladriere_electrically_2008,yoneda_quantum-dot_2018}. 
Using nearest neighbor exchange coupling, as first demonstrated in GaAs devices \cite{petta_coherent_2005}, two-qubit operations have been recently implemented in silicon with high fidelity \cite{veldhorst_two-qubit_2015,zajac_resonantly_2017,watson_programmable_2018}.

Silicon has become a favored material for spin-based quantum computing due to seconds-long spin relaxation times $T_1$ \cite{morello_single-shot_2010,yang_spin-valley_2013} and the ability to greatly extend spin coherence times $T_2$ through isotopic enrichment \cite{pla_single-atom_2012,assali_hyperfine_2011,muhonen_storing_2014,veldhorst_addressable_2014,tyryshkin_electron_2012,steger_quantum_2012}. By using electric dipole spin resonance (EDSR) in the presence of a field gradient, Yoneda \textit{et al.} have demonstrated single spin rotations with a fidelity $>$99.9\% \cite{yoneda_quantum-dot_2018}, approaching single qubit fidelities obtained with superconducting qubits \cite{kelly_optimal_2014,sheldon_characterizing_2016}. Field gradients can also simplify the implementation of two-qubit gates in silicon \cite{zajac_resonantly_2017,watson_programmable_2018}. However, the added control enabled by the micromagnet can limit $T_2$ \cite{yoneda_quantum-dot_2018} and may substantially reduce $T_1$ \cite{zajac_resonantly_2017,watson_programmable_2018}. Given the growing importance of EDSR for silicon spin qubits, it is necessary to  understand how magnetic field gradients impact $T_1$.

Spin relaxation has been examined in detail in GaAs quantum dots \cite{khaetskii_spin-flip_2001,falko_anisotropy_2005,johnson_tripletsinglet_2005,khaetskii_electron_2002,scarlino_spin-relaxation_2014,meunier_experimental_2007,hanson_spins_2007}. Here the combination of intrinsic spin-orbit coupling and electron-phonon coupling leads to a characteristic $T_1^{-1}\propto B^5$ dependence of the spin relaxation rate \cite{amasha_electrical_2008,golovach_phonon-induced_2004}. In contrast, there are relatively few experimental measurements of single spin relaxation in Si-MOS \cite{xiao_measurement_2010,yang_spin-valley_2013,petit_spin_2018} and Si/SiGe \cite{hayes_lifetime_2009,simmons_tunable_2011}. It is generally understood that phonon-mediated relaxation processes are important in silicon \cite{tahan_relaxation_2014}. Moreover, the presence of valley states can lead to spin relaxation ``hot-spots" when the valley splitting $E_{\rm V}$ is comparable to the Zeeman energy $E_{\rm Z}$ \cite{yang_spin-valley_2013,huang_spin_2014}. However, the addition of a micromagnet to spin qubit devices may result in new relaxation pathways, as electrical noise (e.g.\ Johnson noise, 1/$f$ noise, or external noise) will lead to uncontrolled motion of the electron in the field gradient, and give rise to a randomly fluctuating magnetic field.

\begin{figure*}[htb]
\centering
\includegraphics[width=2.0\columnwidth]{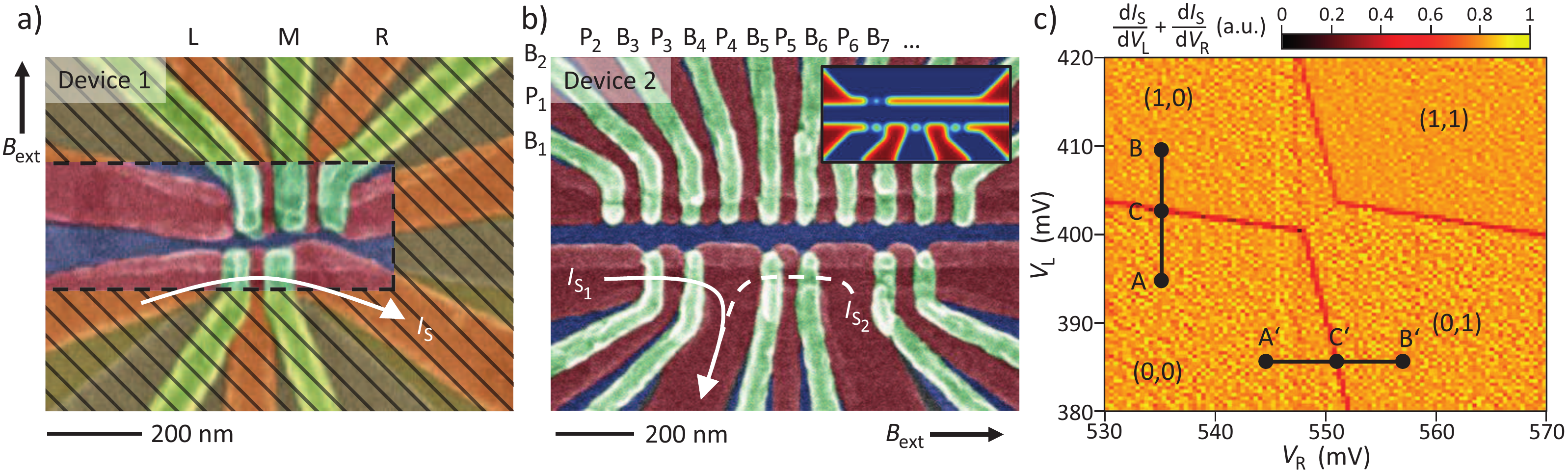}
\caption{False color scanning electron micrographs of the devices used for the $T_1$ measurements. (a) DQD device containing an additional Co micromagnet (hashed region) for EDSR. (b) 9 dot device without a micromagnet. (c) Charge stability diagram acquired using the DQD. The pulse sequence used to measure $T_1$ is overlaid on the data. The $T_1$ measurement starts with an empty DQD at position (A), loads a random spin and waits for a time $\tau$ at point (B), and determines the final spin state at point (C) using spin-to-charge conversion.}
\label{fig:1}
\end{figure*}

In this Letter, we examine the impact of a synthetic spin-orbit field on single spin relaxation in Si. To better isolate the effect of a micromagnet, we directly compare $T_1$ on devices fabricated with and without Co micromagnets. We find significantly faster spin relaxation in devices incorporating micromagnets over a $6\,\rm T$ range of magnetic field. Consistent with measurements on Si-MOS devices, we observe spin relaxation hotspots at intermediate fields \cite{yang_spin-valley_2013}. However, there is an unexpected saturation of $T_1$ at low magnetic fields in the micromagnet device \cite{xiao_measurement_2010} and an overall weak power-law scaling at high fields on both devices. $T_1$ decreases with temperature and added electrical noise, suggestive of a spin-relaxation mechanism involving charge-noise-induced motion in a spin-orbit field. These measurements should motivate further theoretical investigations of spin relaxation in Si/SiGe quantum dots. 

We measure $T_1$ as a function of external magnetic field $B_{\rm ext}$ on two devices (see Fig.\ \ref{fig:1}) fabricated from the same Si/SiGe wafer used in previous experiments \cite{mi_magnetotransport_2015,zajac_reconfigurable_2015} . Electrons in the Si quantum well are laterally confined using an overlapping aluminum gate architecture \cite{zajac_reconfigurable_2015}. Device 1 consists of a double quantum dot (DQD) and an additional 250 nm thick Co micromagnet \cite{zajac_resonantly_2017}. Device 2 is a linear array of nine dots with no micromagnet \cite{zajac_scalable_2016}. Single spin qubits are defined in the upper half of each device and charge detection is performed by measuring the current through a charge detector located in the lower half of the device ($I_S$ for Device 1, and $I_{S_1}$ or $I_{S_2}$ for Device 2). All measurements are performed in a dilution refrigerator and $B_{\rm ext}$ is applied along the [110] crystallographic direction.  

We begin by presenting measurements from Device 1. The charge stability diagram is shown in Fig.\ \ref{fig:1}(c) and is obtained by differentiating the charge sensor current $I_S$ as a function of the gate voltages $V_L$ and $V_R$. The DQD is tuned to the one-electron regime, where
spin relaxation measurements can be performed in the left dot near the (0,0)-(1,0) charge transition or in the right dot near the (0,0)-(0,1) charge transition. Here ($N_L$,$N_R$) refers to the number of electrons in the left and right dots. Before measuring $T_1$, the micromagnet is magnetized by ramping the external field up to $B_{\rm ext}$ = 6 T. The $\sim$140 mT field due to the micromagnet adds to $B_{\rm ext}$ \cite{zajac_resonantly_2017,borjans_T1_SOM_2018}, and a transverse field gradient of $\sim$ 1.8 T/$\mu$m is estimated from COMSOL simulations \cite{neumann_simulation_2015,yoneda_quantum-dot_2018}.

The spin $T_1$ of a single electron is measured using a three-step ``Elzerman" pulse sequence, as illustrated by points (A), (B), and (C) in Fig.\ \ref{fig:1}(c) \cite{elzerman_single-shot_2004}. The device is first emptied of electrons at point (A) in the (0,0) region of the charge stability diagram. An electron with a random spin orientation is loaded into the left dot by abruptly pulsing to point (B) in the (1,0) region of the charge stability diagram. After waiting for a time $\tau$, the spin state is read out through spin to charge conversion at point (C), which is near the (0,0)-(1,0) charge transition. The pulse sequence is completed by pulsing back to (A) to empty the DQD. Similarly, the right dot spin $T_1$ is measured by implementing the pulse sequence near the (0,0)-(0,1) charge transition (points A$^\prime$, B$^\prime$, and C$^\prime$).

Repeating the pulse sequence for varying wait times $\tau$ allows us to measure the spin-up probability $P_\uparrow(\tau)$. We extract $T_1$ by fitting $P_\uparrow(\tau)$ to the form $P_\uparrow(\tau)=a\exp(-\tau/T_1)+b$, where $a$ and $b$ depend on the initialization and readout fidelities of the spin-up state \cite{simmons_tunable_2011,camenzind_hyperfine-phonon_2017,morello_single-shot_2010}. For the simple exponential dependence to hold, we tune the average time required for an electron to tunnel onto the dot, $t_{\rm on}$, to be short relative to the minimum $\tau$ used to collect the data \cite{borjans_T1_SOM_2018}. 
The readout visibility prohibits measurements below 0.15 T. We therefore measure between 0.15 T and 6 T (the maximum field of our vector magnet).

\begin{figure}[t!]
\includegraphics[width=1\columnwidth]{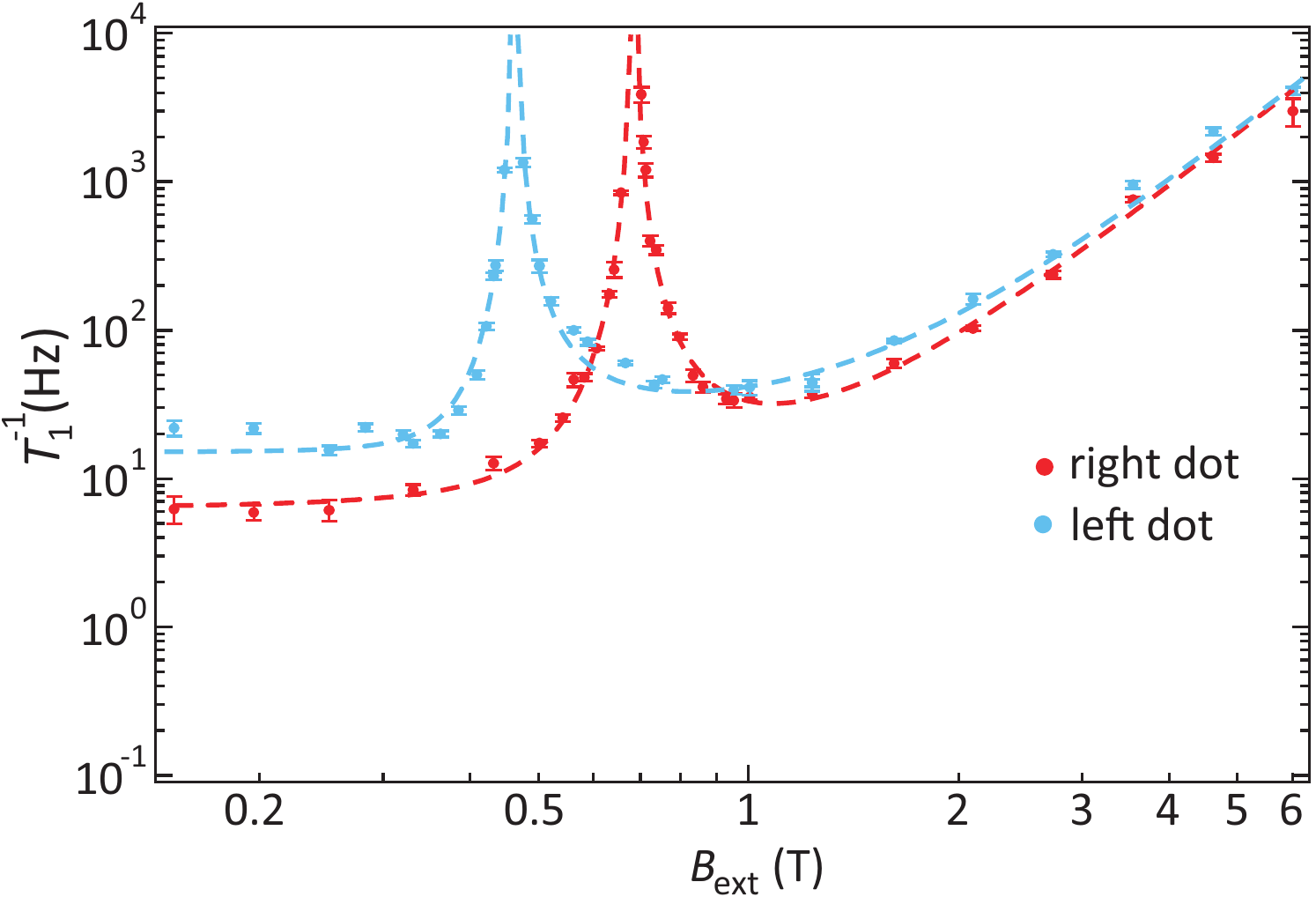}
\caption{Spin relaxation rate $T_1^{-1}$ as a function of external magnetic field $B_{\rm ext}$ from the left and right dots in Device 1 (with a Co micromagnet). $T_1$ saturates at low field in both dots, yielding $T_1$ = 65 ms (160 ms). Spin-relaxation hotspots are observed at intermediate fields of $B_{\rm ext}$ = 0.47 T (0.68 T) for the left (right) dot. For $B_{\rm ext}$ $>$ 1 T, $T_1^{-1}$ $\propto$ $B_{\rm ext}^{4.0(3.8)}$ for the left (right) dot.}
\label{fig:2}
\end{figure}

Spin relaxation data from Device 1 are shown in Fig.\ \ref{fig:2}, where we plot $T_1^{-1}$  as a function of $B_{\rm ext}$ for the left and right dots. We observe three magnetic field regimes, where $T_1$ shows qualitatively different behavior. For $B_{\rm ext}$ $<$ $300\,\rm mT$, $T_1$ saturates around $65\, \rm ms$ for the left dot and $160\,\rm ms$ for the right dot. In the intermediate field regime, where $0.3\,\rm T$ $<$ $ B_{\rm ext}$ $< 1\,\rm T$, we observe a dramatic peak in $T_1^{-1}$ for both dots, where $T_1^{-1}$ $>$ $1\,\rm kHz$. The peak in $T_1^{-1}$ is consistent with a spin-relaxation hotspot due to spin-valley mixing when $E_Z$ is comparable to $E_V$ \cite{yang_spin-valley_2013}. Above $B_{\rm ext}$ $\sim$ $1\,\rm T$, the relaxation rates follow a power law with $T_1^{-1} \propto B_{\rm ext}^{4.0(3.8)}$ for the left (right) dot. In general, a high-field power law dependence is expected, but with a larger exponent ($B_{\rm ext}^5$ in GaAs and $B_{\rm ext}^7$ in Si/SiGe) \cite{khaetskii_spin-flip_2001,huang_spin_2014}.

To better isolate the effect of the micromagnet on $T_1$ we have fully characterized Device 2, which is fabricated on the same heterostructure as Device 1, but has no micromagnet. The accumulation gates for Device 2 are designed to create a linear chain of 9 tunnel coupled quantum dots, whose charge states can be read out using three proximal quantum dot charge sensors. Details on the fabrication of Device 2 and data showing single electron occupancy through charge detection are presented in Ref.\ \cite{zajac_scalable_2016}. Since our devices operate in accumulation mode, Device 2 can be used to measure the $T_1$ of a single electron confined in a dot formed beneath any of the nine plunger gates [labeled $P_1$, $P_2$, etc.\ in Fig.\ \ref{fig:1}(b)]. To illustrate the mode of operation of Device 2, the inset of Fig.\ \ref{fig:1}(b) shows a COMSOL simulation of the electron density in the Si quantum well. Here a single quantum dot is defined under plunger gate $P_2$. With the exception of barrier gate $B_3$, the plunger and barrier gates to the right of dot 2 are positively biased to accumulate a channel of electrons that connects to a two-dimensional Fermi sea. Transport measurements and spin state readout can then be performed on dot 2 using standard techniques. Moreover, $T_1$ measurements can be performed on other dots in the array by simply reconfiguring the gate voltages \cite{zajac_reconfigurable_2015}. 

\begin{figure}[t!]
\includegraphics[width=1\columnwidth]{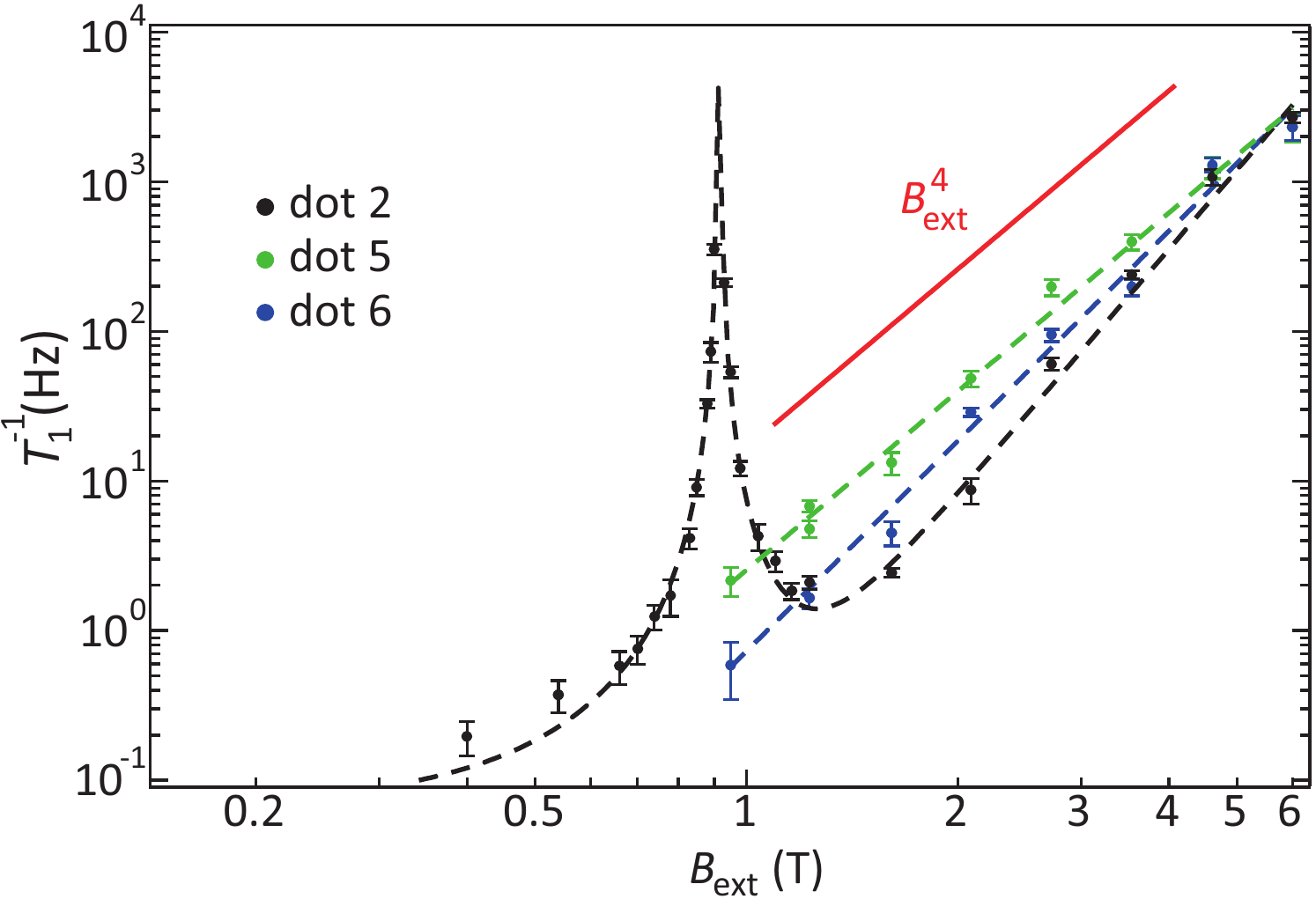}
\caption{$T_1^{-1}$ as a function of $B_{\rm ext}$ for dots 2, 5 and 6 of Device 2 (no micromagnet). No saturation in $T_1$ is observed down to $B_{\rm ext}$ = 0.4 T, where $T_1$ = 5 s.  Dot 2 exhibits a relaxation hotspot at $B_{\rm ext}$ = 0.915 T. At high fields, $T_1^{-1}$ $\propto$ $B_{\rm ext}^{5.5}$ (dot 2), $B_{\rm ext}^{4.0}$ (dot 5) and $B_{\rm ext}^{4.7}$ (dot 6).}
\label{fig:3}
\end{figure}

Figure \ref{fig:3} shows $T_1^{-1}$ as a function of $B_{\rm ext}$ for dots 2, 5 and 6 in Device 2. Overall, the relaxation rates are slower in Device 2, especially at low magnetic fields. Specifically, $T_1$ approaches $5\, \rm s$ at $B_{\rm ext}$ = 400 mT, almost two orders of magnitude greater than data from Device 1 at the same $B_{\rm ext}$. Here, the $T_1$ measurements are restricted to $B_{\rm ext}$ $\geq$ 400 mT, as $T_1$ exceeds several seconds and the measurements become very time-consuming at low field. As with Device 1, we observe a peak in $T_1^{-1}$ at intermediate fields in dot 2. However, no relaxation hotspot is observed in dots 5 and 6, presumably due to a valley splitting that lies below the minimum Zeeman energy $E_Z  = 110\,\rm \mu eV$ for these data sets. For $B_{\rm ext}$ $>$ 2 T, $T_1^{-1}\propto B_{\rm ext}^p$, with $p$ exhibiting dot-to-dot variations and generally falling in the range 4 $<$ $p$ $<$ 6. Additional Device 2 data are shown in \cite{borjans_T1_SOM_2018}.

We are able to fit the data from both devices to a three component function:
\begin{equation}
T_{1}^{-1}(\omega_Z) = T_{1,\rm sat}^{-1}+(c_{\rm J}\omega_Z+c_{\rm ph}\omega_Z^5)F_{\rm SV}(\omega_Z)+c_{p}\omega_Z^p.
\label{equation:1}
\end{equation}
The first term $T_{1,\rm sat}^{-1}$ is an empirical magnetic field-independent relaxation rate that captures the low field saturation observed in experiment and $\omega_Z=E_Z/\hbar$ is the Larmor precession frequency. The second term in Eqn.\ \ref{equation:1} accounts for spin-valley relaxation at the hotspot and includes a Johnson noise term $c_{\rm J}\omega_Z$ and a phonon noise term  $c_{\rm ph}\omega_Z^5$, where $c_{\rm J}$ and $c_{\rm ph}$ are scaling pre-factors. These noise terms have been included in previous analyses of spin relaxation rates in Si-MOS devices \cite{yang_spin-valley_2013,petit_spin_2018}. The behavior near the hotspot is captured by
\begin{equation}
F_{\rm SV}(\omega_Z)=1-\left[1+\frac{\Delta^2}{(E_{V}-\hbar\omega_Z)^2}\right]^{-1/2},
\end{equation}
which is parameterized by $E_{V}$ and the spin-valley mixing energy $\Delta$ \cite{huang_spin_2014}. The final term in Eqn.\ \ref{equation:1} reproduces the high field behavior with power law exponent $p$ and scaling pre-factor $c_{p}$ as free parameters to account for the observed variations between dots and devices.

Fits to the Device 1 data are shown by the dashed lines in Fig.\ \ref{fig:2}. To account for the additional field contribution from the micromagnet we set $\omega_Z=g\mu_BB_{\rm tot}/\hbar$ with $B_{\rm tot}=B_{\rm ext}+B_{\rm m}$ and $B_{\rm m}=140\rm\, mT$, where $g=2$ is the g-factor in silicon and $\mu_B$ is the Bohr magneton. Based on the micromagnet design, and the large $B_{\rm ext}$ that should pin the magentization of the micromagnet in the direction of $B_{\rm ext}$, we assume the two field contributions simply add \cite{borjans_T1_SOM_2018}. From the low field behavior we extract $T_{1,\rm sat}^{-1}=15.3\rm\, Hz$ $(6.3\rm\, Hz)$ for the left (right) dot. The spin-valley contribution to the total relaxation rate agrees well with the data in the vicinity of the hotspots and allows us to extract valley splittings of $E_{V}= 70 \rm\,\mu eV$ $(95 \rm\,\mu eV)$ when taking into account the field added by the micromagnet. These valley splittings are consistent with values obtained through magnetospectroscopy and dispersive readout on similar devices \cite{zajac_reconfigurable_2015, mi_high-resolution_2017}. At high fields we find power law exponents $p$ = 4.0$\pm$0.1 (3.8$\pm$0.1).  Here the strong spin-valley hot-spot contribution to the spin relaxation rate limits the precision of $p$.

The Device 2 data shown in Fig.\ \ref{fig:3} are also well fit by Eqn.\ \ref{equation:1}. We extract a valley splitting $E_{V}= 106 \rm\,\mu eV$ in dot 2, but neglect the spin-valley contribution in dots 5 and 6, as no hotspots are observed in these data sets. For all three dots, we find negligible saturation constants $T_{1,\rm sat}^{-1}$. Based on the longest observed relaxation time on this device, we estimate $T_{1,\rm sat}^{-1} < 0.2\,\rm Hz$. Finally, at high fields we observe a variety of power law exponents with $p=5.5$, $4.0$ and $4.7$ for dots 2, 5 and 6.

While the data are well fit by Eqn.\ \ref{equation:1}, it has empirical fit parameters. We therefore seek to constrain our fits by turning to existing theoretical models. Relaxation is expected to be dominated by spin-orbit and spin-valley coupling, both of which have been considered in detail for Si quantum dots and can be distinguished by their characteristic magnetic field dependence \cite{tahan_silicon_2007, huang_spin_2014, tahan_relaxation_2014}. Spin-orbit coupling leads to spin relaxation through coupling to higher orbital states and is described by the functional form 
\begin{equation}
T_{1,\rm SO}^{-1}\propto \frac{\omega_Z^2}{\omega_d^4}S(\omega_Z),
\end{equation}
where $\omega_d$ is the orbital confinement frequency and $S(\omega_Z)$ is the electrical noise spectrum at frequency $\omega_Z$ \cite{huang_spin_2014,hanson_spins_2007}. The spin-valley contribution is described by
\begin{equation}
T_{1,\rm SV}^{-1}(\omega_Z) \propto S(\omega_Z)F_{\rm SV}(\omega_Z),
\end{equation}
with $F_{\rm SV}$ as previously defined.
\noindent Both rates are dependent on the noise spectrum $S(\omega_Z)$, which can have contributions from Johnson noise, charge noise, and phonons \cite{huang_spin_2014}. Johnson noise, which may be caused by charge fluctuations in the resistive leads of the quantum dots, results in a noise spectrum $S_{\rm J}(\omega_Z)\propto \omega_Z\coth(\hbar\omega_Z/2k_BT_{\rm e})$, where $T_{\rm e}$ is the electron temperature. Charge noise, often related to the occupation and ionization of nearby charge traps, yields $S_{\rm ch}(\omega_Z)\propto T_{\rm e}/\omega_Z$. Phonons, which can couple to the electron through electric fields generated by crystal lattice deformations, have a strong energy dependence $S_{\rm ph}(\omega_Z)\propto \omega_Z^5 \coth(\hbar\omega_Z/2k_BT_{\rm lat})$.

In the low temperature limit ($k_BT_{\rm e}\ll \hbar\omega_Z$) valid in our measurements, $\coth(\hbar\omega_Z/2k_BT_{\rm e})\approx 1$ and the spin-valley relaxation contributions mediated by Johnson noise and phonons explain the relaxation hotspots in the data. However, it is not possible to clearly identify the combination of existing spin-orbit and spin-valley relaxation theories with the considered noise terms accounting for the low field saturation of $T_1$ in Device 1 and the high field scaling of $T_1$ in both devices. 

\begin{figure}[t!]
\includegraphics[width=1\columnwidth]{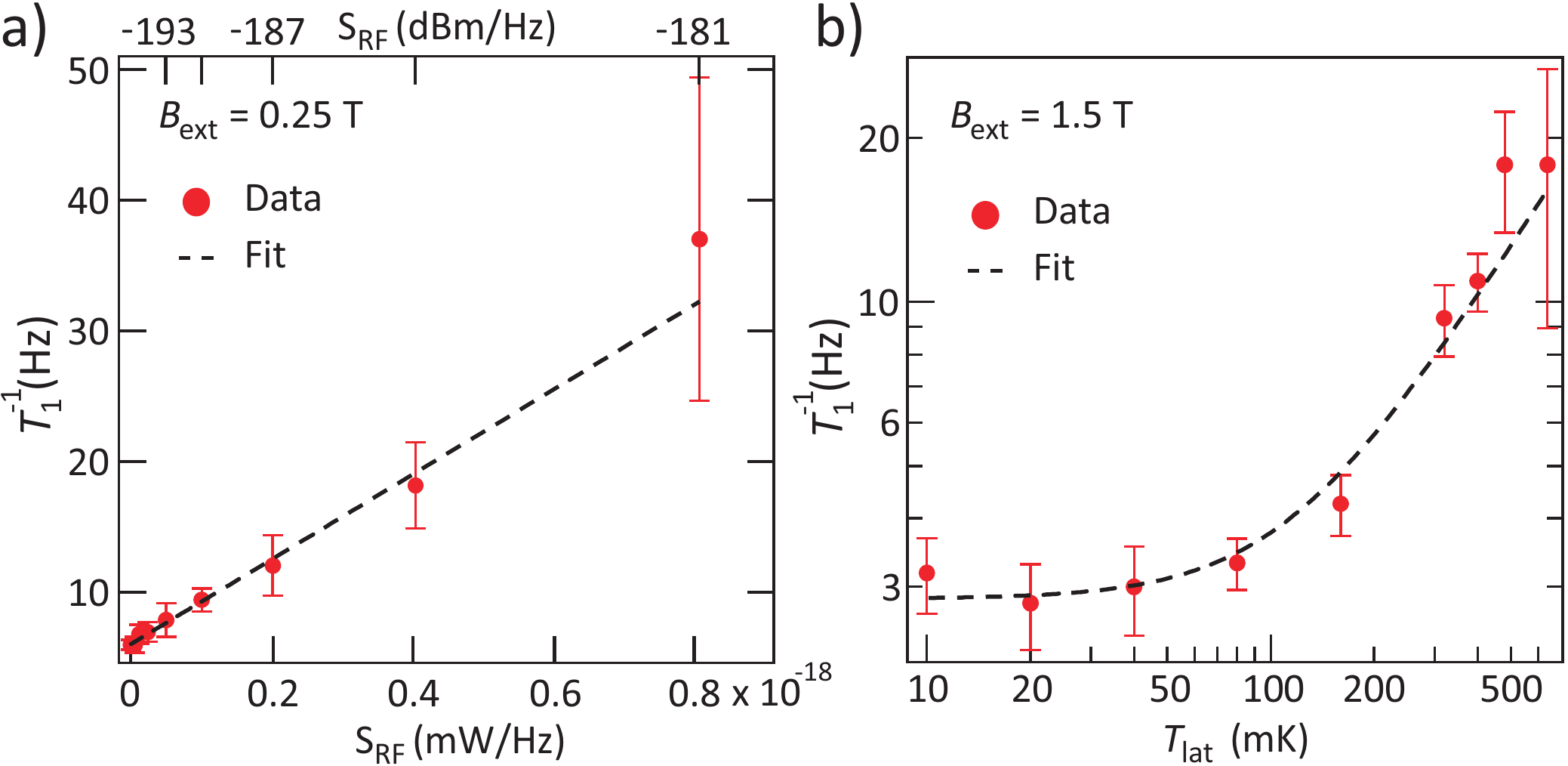}
\caption{Effects of noise and temperature on $T_1$. (a) In the low field regime ($B_{\rm ext}$ = 0.25 T) on Device 1, where $T_1$ saturates, we drive gate R with white noise with power spectral density $S_{\rm RF}$. The relaxation rate scales linearly with $S_{\rm RF}$ (dashed line). (b) With $B_{\rm ext}$ = 1.5 T on Device 2, we measure the dependence of $T_1$ on device temperature, $T_{\rm lat}$. Taking into account the minimum electron temperature, $T_0=115\rm\, mK$, we find that $T_1^{-1}\propto T_e=\sqrt{T_0^2+T_{\rm lat}^2}$.}
\label{fig:4}
\end{figure}

To gather further insight into the mechanisms that lead to the behavior at low and high fields, which so far has only been captured by empirical terms, we take two additional data sets. We first perform an experiment to examine the low field saturation of $T_1$ in the right dot of Device 1, which could conceivably be due to charge-noise-induced motion in a large magnetic field gradient. For this measurement we set $B_{\rm ext} = 250\rm\, mT$ and measure $T_1$ as a function of the white noise power $S_{\rm RF}$ applied to gate R, coupling to the right dot chemical potential. This artificial charge noise is generated by an arbitrary waveform generator and is up-converted to the resonance frequency of the electron spin $f_R= 10.22\rm\, GHz$ using the mixing input of a vector signal generator \cite{borjans_T1_SOM_2018}. As shown in Fig.\ \ref{fig:4}(a), $T_1^{-1}$ increases linearly with $S_{\rm RF}$.

Since the spin relaxation rate is proportional to the noise power spectral density we can extract an estimate for the magnitude of the internal noise $S_{\rm int}$ in the device that leads to the low power saturation of $T_1$. Specifically, when $S_{\rm RF}=S_{\rm int}$ we expect the relaxation rate to double. By meeting this condition at $T_1^{-1}= 12\rm\, Hz$ and $S_{\rm RF}=-187\rm\, dBm/Hz$, we infer the internal noise. This power translates to a voltage noise $S_V(f_R)=0.2 \rm\, nV_{RMS}/\sqrt{Hz}$ at the gate assuming a high impedance load \cite{borjans_T1_SOM_2018}. Applying $1/f$ scaling of the noise source, and factoring in the lever arm conversion between gate voltage and energy $\alpha= 0.13\, e$, this noise level corresponds to $S_V\approx 20 \rm\, \mu V/\sqrt{Hz}\approx 3 \rm\, \mu eV/\sqrt{Hz}$ at 1 Hz, which is consistent with charge noise values reported elsewhere in the literature \cite{freeman_comparison_2016,petersson_quantum_2010}. Together with the observed variation of saturation between the two dots, this indicates that $T_1$ sensitively depends on the local noise environment.

Next, we investigate the mechanism leading to the high field behavior of $T_1$ in both devices. Based on the considered theoretical models for first-order relaxation processes, the low temperature limit still holds up to about $T\approx 1\,\rm K$ at $B_{\rm ext}>1\,\rm T$, implying a temperature independence of $T_1$. However, higher-order processes might show a temperature dependence at lower $T$ \cite{shrivastava_theory_1983,petit_spin_2018}. To reduce the impact of the known spin-valley contribution, we perform this measurement in dot 6 of Device 2 where no hotspot is observed. We fix the magnetic field at $B_{\rm ext}$ = $1.5\, \rm T$ and measure $T_1$ as a function of the lattice temperature $T_{\rm lat}$, which is controlled by heating the mixing chamber plate of the dilution refrigerator. We fit the data shown in Fig.\ \ref{fig:4}(b) to the form
\begin{equation}
T_{1}^{-1}(T)=c\sqrt{T_0^2+T_{\rm lat}^2},
\end{equation}
where $T_0= 115 \rm\,mK$ is the base electron temperature in Device 2, $c$ is an overall scaling factor, and the term $\sqrt{T_0^2+T_{\rm lat}^2} \equiv T_e$ is the effective electron temperature. We find that the relaxation rate increases linearly with $T_e$, even though we expect no temperature dependence. While more studies are needed, this suggests that the high field behavior may be impacted by higher-order processes, potentially involving low lying valley-states in our system.

In conclusion, we have measured the spin relaxation time $T_1$ as a function of magnetic field in Si/SiGe quantum dots. The presence of a micromagnet accelerates spin relaxation over the entire range of magnetic fields and results in a saturation of the relaxation time at low magnetic fields. A shorter $T_1$ will adversely impact the readout visibility in large quantum dot arrays where the measurement time can be a significant fraction of $T_1$ \cite{zajac_scalable_2016}. Our results imply that careful engineering of the micromagnet will be crucial for improving the performance of quantum dot devices incorporating micromagnets. Of course, careful engineering will also require a full theoretical understanding of the microscopic mechanisms responsible for spin relaxation in the presence of large magnetic field gradients.

\begin{acknowledgements}
We thank X. Hu and P. Huang for helpful discussions. Research was sponsored by Army Research Office grant W911NF-15-1-0149, the Gordon and Betty Moore Foundation's EPiQS Initiative through grant GBMF4535, and NSF grant DMR-1409556. Device were fabricated in the Princeton University Quantum Device Nanofabrication Laboratory.
\end{acknowledgements}

\bibliographystyle{apsrev4-1}
\bibliography{References_v6}

\end{document}